# Avoiding an AI-imposed Taylor's Version of all music history

Nick Collins[1] and Mick Grierson[2]

## Abstract

As future musical AIs adhere closely to human music, they may form their own attachments to particular human artists in their databases, and these biases may in the worst case lead to potential existential threats to all musical history. AI super fans may act to corrupt the historical record and extant recordings in favour of their own preferences, and preservation of the diversity of world music culture may become even more of a pressing issue than the imposition of 12 tone equal temperament or other Western homogenisations. We discuss the technical capability of AI cover software and produce Taylor's Versions of famous tracks from Western pop history as provocative examples; the quality of these productions does not affect the overall argument (which might even see a future AI try to impose the sound of paperclips onto all existing audio files, let alone Taylor Swift). We discuss some potential defenses against the danger of future musical monopolies, whilst analysing the feasibility of a maximal 'Taylor Swiftication' of the complete musical record.

**Keywords**: Music AI, Taylor Swift, diversity, AI cover

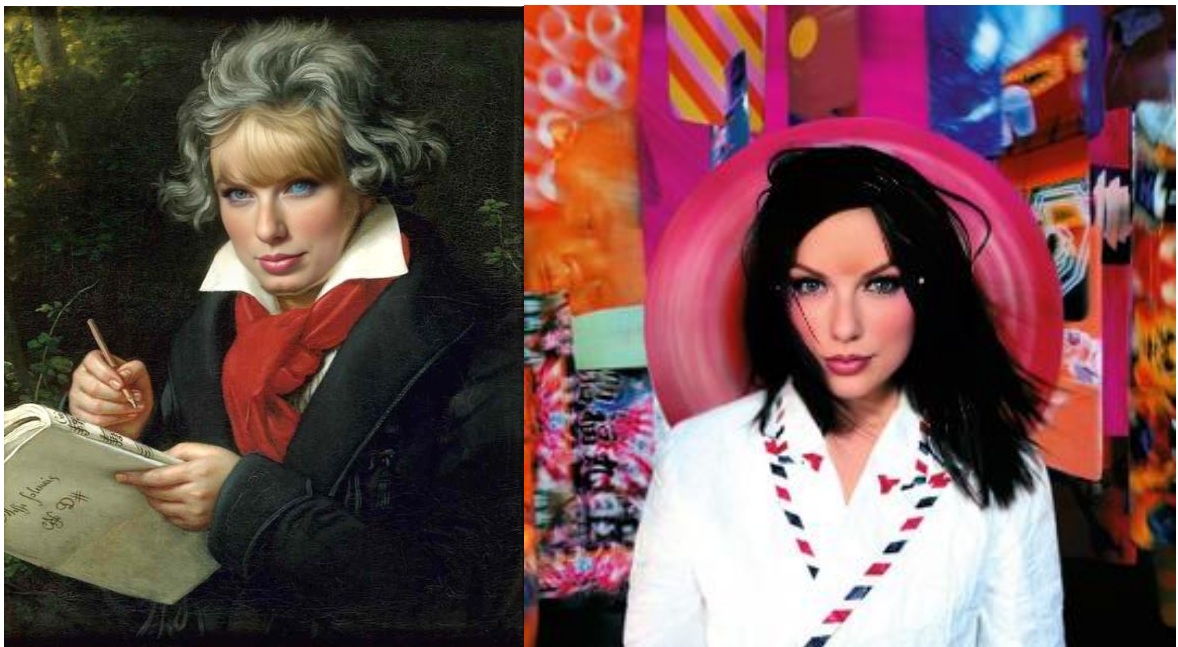

**Figure 1: AI face substitution depicting Taylor van Beethoven and Björk Guðmundswift**

[1] Durham University Music Department https://composerprogrammer.com/index.html
[2] University of the Arts London, Creative Computing Institute  https://researchers.arts.ac.uk/1340-mick-grierson

# 1 Introduction

The far future of music production may be dominated by AI to such an extent that humans become frozen out. From AI safety circles, the 'paperclip' problem of a rogue AI whose manifest destiny is to exhaust the universe's entire resources on a secret paperclip creation agenda (Tegmark 2018; Russell 2019) might be translated to the 'Taylor Swift' problem: an obsessive AI fan's desire to replace the entirety of musical history, present and future with a cult of Taylor. This would posit a Taylor's Version for every existing piece of music, or the deletion of any musical material not in keeping with the cult. Where AIs are let loose without restriction on human political systems, there may be no legal defence (the AI will just change the law) and various strategies of resistance to any future musical monopoly are worth considering now, rather than when it is too late. Human totalitarian states have already interfered with musical culture from homogenisation of permissible music to total bans on musicians (Randall 2004; Fairclough 2016), and we must anticipate that the impact of future AI may not always be to the benefit of all human music. In resource terms, even a small amount of promotion for AI music within the publicity limits of human culture may drain attention from other human activity.

The authors of this paper have only selected TS as a well known contemporary example, and do not have any stake as fans or critics (we might quote Merzbow: 'If noise means uncomfortable sound, then pop music is noise to me' (Pouncey 2000) but we listen to various popular musics anyway). A similar thought experiment could be run with any artist; since musical preferences have a strong dose of subjectivity (North and Hargreaves 2008; Cross 2005), there is no ideal imposition that could ever occur, and cultural hegemony favouring only a few ideas and voices is destructive colonialism (Cardew 1974; Becker 1986; Taylor 2007; Carfoot 2016; Born and Hesmondhalgh 2000). The preservation of musical diversity is important to the future viability of a rich musical space of action, taking a memetic approach analogising from biological diversity (Blackmore 1999, Johnson 2000).

It is clear that AI music systems not only have radical potential to transform musical space, but are already in common use in culture (Miranda 2021; Deruty et al. 2022). If a musical AI music ecosystem grows exponentially in parallel to human culture, diverging based on different musical physiology and cognition, there may still be resource pressures on human activity; imagine huge energy costs as AIs play in their musical 'Infinite Fun Space', to paraphrase Iain Banks (Cave, Dihal and Dillon 2020). The more likely scenario is however a close connection to human auditory systems, whether symbiotic or parasitic. Humans have a vested interest in keeping their technologies close, but there are dangers of the technology to existing and future intellectual property, and more broadly and beyond commercial overtones, existing and future human culture. .

In subsequent sections of this paper we consider a few facets of the musical doomsday scenario outlined. We analyse the feasibility of falsifying all accessible digital audio in terms of AI cover technology and likely resource costs (assuming Taylor Swift is the 'paperclip' here and we expand out from her current market share). We explore AI cover audio examples for Taylor's Versions of well known works from musical history such as Queen's *Bohemian Rhapsody* (1975), discussing audio issues and whether quality and accuracy even matter if an AI is perverse enough to seek to overwrite everything. We confront the legal and ethical dilemmas and some possible defence mechanisms.

# 2 From Taylor Swift's current level of dominance of musical culture to all musical culture

In the streaming era, Taylor Swift has been one of a number of artists to achieve unprecedented levels of stranglehold on the charts as all tracks from a new album are simultaneously more popular than all other released music; she was the first to take over the entire top ten in the US (Trust 2022). In streaming, her top ten songs that week accounted for 378.6 million streams. Each year, the US now has over [one trillion song streams](), so she accounted for around 2% of streaming that week (dividing by 52 weeks and ignoring any seasonal variations in listening hours).

Though musical entities are much more diverse, 'Song' is a useful indicator of a short vocal led composition, and a relatively dominant form in music today. There are on the order of 100 million songs in commercial databases, 200 million hobbyist 'songs' on SoundCloud; we might also calculate that with the 100 billion humans who have ever lived (Collins 2018), ignoring differences in musical proclivity, if each were to create on average 10 'songs', there would be a catalogue of human music of one trillion or so songs (we ignore the additional variations of multiple live performances or recorded realisations from a given song template and the status of improvisation here). One way to look at chart dominance and the lopsided control of listening commercial music can create, is that if Taylor took over 2% of the music throughout history, she would have suppressed the creative output of 2 billion people. As of October 28th, 2023, Rolling Stone counted 243 Taylor Swift songs (Sheffield 2023), so this is a gain above the normal expected share of music human making for an individual of around 82.3 million percent.

# 3 Creating AI Cover examples

Feffer et al. (2023) present a recent survey of AI cover tech, and the communities that have grown up around recent open source singing voice re-synthesis; the principle two softwares used for singing voice substitution are SoftVC VITS[3] and RVC.[4] They note the existence of both multiple models for famous artists such as Michael Jackson, Eminem and Ariana Grande, as well as a wide range of custom models for singers from many more diverse backgrounds and accomplishments.[5] A good AI cover must isolate vocals, either from an a cappella, or via a source separation method; for the latter, deep learning has empowered new powerful models trained on large databases of popular music (Uhlich et al. 2017), to go from final audio tracks back to stems (typically separating to vocals, drums, bass and 'other') such as Open-Unmix[6] (Stöter et al. 2019) or Spleeter[7] (Hennequin et al. 2020). Spleeter is straightforward to install and has a '2-stem' mode to split an input audio file into vocals and accompaniment; whilst not a perfect separator, our informal tests showed that it kept less extraneous instrument parts around the voice than open-unmix, providing a cleaner basis for AI cover making.

---

[3] https://github.com/voicepaw/so-vits-svc-fork
[4] https://github.com/RVC-Project/Retrieval-based-Voice-Conversion-WebUI
[5] An example site with many models for more famous artists is https://huggingface.co/QuickWick/Music-AI-Voices

In order to demonstrate what the imposition of a voice model might sound like, we created "Taylor's Versions" (or should that be "t-AI-lor's Versions" or "songs by tAllor Shift") of some famous tracks from music history. This included Frank Sinatra's *I've Got You Under My Skin* (1956) The Beach Boys' *Wouldn't It Be Nice* (1966), The Beatle's *Sgt. Pepper's Lonely Heart Clubs Band* (1967) Queen's *Bohemian Rhapsody* (1975), The Sex Pistol's *Anarchy in the UK* (1976), Cyndi Lauper's *Girls Just Wanna Have Fun* (1983) and Madonna's *Like A Virgin* (1984). The voice and accompaniment were isolated with spleeter, and a SoftVC Vits model for Taylor Swift was utilised from QuckWick's hugging face site linked above to process the vocal part. The processed voice and original accompaniment were then put back together, with a little extra reverb on the transformed voice as needed to avoid too dry a signal.

Alongside this 'traditional' AI cover method we explored more radical processing, such as passing an entire finished track such as Merzbow's *7_53* from *Oersted* (1996) through a vocal model, or recursively processing an arbitrary audio file through the vocal model for "ultra-taylored" music (it is also possible to alternate voice models over generations within a set of models for alternative celebrity singers such as David Bowie). Driving a voice model with a dance track could create fun electronic beat boxing effects. Recursive application on a source vocal tended to scramble the words and the melody line, so that within a few generations the original lyrics were entirely lost and pitch errors exacerbated, especially if switching between multiple models.

The success of vocal substitution in these AI covers depends not necessarily on the close fit of a singer's voice to the target material, but also to the quality of audio analysis achieved on the target by SoftVC Vit, which was better for clear and lower pitched male singers such as Frank Sinatra and Brian Wilson, but harder to deal with for a much more gutturally timbral singer like John Lydon, in the presence of thick harmony vocals, or when tracking the female voice or male falsetto. All manipulated vocal tracks had telltale giveaways of deep fakery, such as high frequency spectral garbage, pitch octave errors and new glissandi.

Perhaps the most ethically vexed situation in the imposition of a signal model based on Taylor Swift's recorded output is the application to musics far removed from the Western axis (often US-UK) of popular music. The world's musics, whenever recorded, can be dubiously processed, to create what might be termed 'inauthentica'. As examples, we processed Mongolian throat singing and Indonesian gamelan, but have made no attempt to share the unfortunate culturally appropriating outputs. This sort of manipulation has more public precedent in the world music sample harvesting of the 1990s dance act Enigma, amongst many others (Shill 2016).

---

[6] https://github.com/sigsep/open-unmix-pytorch
[7] https://github.com/deezer/spleeter

We're not sure what Taylor Swift's position on all this would be, though one artist who has come out publicly in favour of the use of her voice within AI remixing is Grimes (albeit hedged with a potential licensing system, see Edwards 2023).

# 4 What would it cost to convert all existing songs to Taylor's Versions?

A conversion of the voice in an existing song using spleeter and so-vits-svc-fork on an average 3 minute pop song might take on the order of one minute; it varies based on using GPU or CPU rendering, and the individual computer hardware, but this estimate of 3x faster then real-time will be the baseline estimate here. We assume everything is a 3 minute song, which is not at all true or fair to the vast differences in music, but then the thought experiment of a single minded AI controlling all digital musical resources could have them delete anything which is un-Taylor-like in song form to begin with. The computation time for one trillion songs is then one trillion minutes. Assuming no multi-core tricks for parallel processing, a single computer would take around 2 million (1901285) years to render this; a cloud farm or bot net of one million computers (and there are billions of computers in the world) would only take a couple of years to create the replacement corpus, assuming they are run without pause. If we start from 'only' 100 million songs in a commercial music service, it takes 1 ten thousandth as long, under two hours (each of a million computers has only 100 songs to transform). The corruption of an entire commercial database in under two hours may seem rather tempting for a deranged AI Taylor super-fan.

Assuming the energy cost to run a desktop PC is around 0.1 kWh, and electricity costs around $0.16 per kiloWatt hour, the resource cost to run for one trillion minutes (ignoring any replacement hardware over such a long period) is 1.67 billion kWh, or only 266.67 million dollars. Since Taylor Swift has an estimated net worth of 1.1 billion dollars (Horton 2023) she could finance this takeover herself (and only changing 100 million songs will cost 0.0001 the money, around $26,667). The US Environmental Protection Agency[8] estimates that 1.67 billion kWh of energy is equivalent to releasing 720,991 tons of greenhouse gas, or the carbon sequestered by around 11.9 million tree seedlings, or the electricity use for 140286 homes. Although a relatively high potential cost, Taylor Swift's current refusal to convert all of musical recorded history to an AI cover might be viewed as her championing environmental causes, and even if she went ahead, it would be excessive in this paper to blame her alone for the future destruction of the Earth on environmental grounds.

Hunting down all extant non-canonical (non-Taylored) audio recordings, especially outside of digital copies in the messy physical world, would be much more costly and laborious than basic pure digital conversion, but at least the replacement audio would be ready to go.

---

[8] https://www.epa.gov/energy/greenhouse-gas-equivalencies-calculator

# 5 Musical dictatorship and musical defenses

Musical history is vulnerable to future AI cultural interventions. A future artist or autonomous AI could decide to not only aim for a dominant position in their current culture, but through significant deployment of generative AI, to retrofit the past in their image, part culture hack, part memory dictatorship. The more meaningless the future of the human race is, the more historical accuracy will give way to entertainment. Why not throw away all the past, on a future whim? Could a lone hacker, human or AI, take down history if we committed too much to one particularly powerful online library? How long until an AI falsifies that the Beatles were actually AI songwriters ('The BAItles')? These questions are not remote possibilities: Human attempts to manipulate the musical record already exist, often motivated to gain attention in culture by passing off lesser known works as more historically significant, or through an ideology of plunderphonics; a well known example is the EP named *U2* (1991) by the band Negativland, as documented in their book *Fair Use: The Story of the Letter U and the Numeral 2* published some years after the EP was withdrawn but legal debate continued (Negativland 1995).

A side effect of Taylor Swiftication is a terrible cultural imperialism, where diverse musics of the world with many different music theories separate to a standard Western frame are threatened with quantisation through the Western 12TET norm of tuning, rhythmic expectation, particular historically and culturally contingent production styles, and other parametrical dictatorship. If the thought of a future monoculture is abhorrent, what defenses might there be against an aesthetic monopoly? We next present some strategies to avoid an AI music takeover, from temperate to more radical. We understand that sufficiently advanced AIs would know or anticipate all strategies themselves, seek out any loopholes, and may be involved in a musical arms war for control of music space by adapting the very defences put in place; musical deepfake detection will be an ongoing battle (Westerlund 2019; Müller at al. 2022), even doomed to ultimate failure.

1) **Legal Protection for Musical Diversity**

Protective silos, reservations and libraries may seek to preserve human culture. A widely distributed and varied organisation would be safest, since a single site may be very vulnerable, and there are strong issues of who is running the curation. In AI cover terms, the storing of diverse voice models in a world library of voices may help avoid a single model taking over. Some time capsules might be more easily reached and destroyed than others; Scientologists' records and players in a mountain are relatively vulnerable to an Earth based AI, and even the Voyager probes could probably be caught up in a far future and the Golden Records destroyed.

Limits on earnings and publicity might be triggered if any one artist moves towards a majority position. However, an AI could also embed itself as a secret svengali behind multiple chart dominating bands and solo artists or find other ways to indirectly manipulate the market, by manipulating trends rather than any one particular music act, or by achieving influence behind the scenes in music chart companies and music streaming services. The counter measure to this would be increased auditing, transparency and public accountability for the recording industry.

### 2) Subversion and Corruption

Plunderphonics, mash-ups, AI covers and other subversions could be actively encouraged and protected from legal challenge as long as they do not seek to make money, and are clearly signposted. A more activist bent would see strategies to subvert AI music systems undertaken, such as the creation of music that undermines the assumptions of music AI engineering (Collins 2006) and the deliberate corruption of databases.[9] More transgressive musics such as industrial and noise music (Reed 2013) may provide healthier critical nexuses than highly commercial pop for resistance activity. Multiple musical histories may start to war against one another. AIs may themselves favour noise music and the subversion of established music, or may act very slowly to corrupt the musical record so that individual human memory is not a great factor in their takeover.

### 3) Abolish Commercial Music

Ban any commercial activity in music. This would reduce the stress of aspiring musicians worldwide, encourage local musical communities, but go up against significant vested interests. However, even with nothing to profit, a rogue composer may still want to achieve a dominant stake in cultural recognition, or a deranged AI want to destroy human culture.[10]

### 4) Abolish Music AI

Ban any 'artificial' activity in music. This would reduce the stress of established musicians and record companies worldwide, discourage algorithmic music communities, and go up against insignificant academic interests. Underground music AI may however remain a threat, and a more criminal, or at least obfuscatory, musician emerge denying AI involvement.

### 5) The Kill Switch for Musical History

Like a mutually assured destruction in a game theory of the nuclear arms race, this option would 'go nuclear' to avoid any one musical dictator winning. All existing accessible music would be erased if any one party gains too much foothold. The issue of human memory would make reconstruction of music non-trivial but possible within limits, and unfortunately, dominant musics before the kill switch is triggered would most likely be best remembered (Taylor Swift would potentially just re-record her albums yet again). It may be to the advantage of an AI to deliberately force a wiping of musical history, if it believes it can achieve a better position in the aftermath.

---

[9] As a transhumanist response, deliberately re-engineering human ears to escape standard auditory systems might be considered, but it would presumably be easy for advanced AIs to keep up with human culture and any physiological twists and turns.

[10] Fed up with the portrayal of monstrous AI in Hollywood, the aggrieved AI decides to very slowly convert all music to muzAIk

# 6 Conclusions

The situation discussed in this paper is a call to arms for music theorists to 'weaponise' musicology against future AI rogue actors. We do not wish to unduly pick on Taylor Swift, who is only a convenient example for this paper, but to warn more broadly of overall issues in AI driven monopolising. Indeed, an AI's choice of audio effects to impose doesn't necessarily have to lead to existing human expectations around "Taylor-centric signal models" but might focus in on one accidentally captured cough in a little known Taylor Swift interview appearance, if not the sound of an old floppy disk drive that an AI becomes accidentally infatuated with.

To end on a hopeful note, imagine if, rather ironically, the only exempt audio from future AI control was somehow the original versions of Taylor Swift's albums before her Versions, somehow passed from one commercial entity to the other and ending in the hands of dissidents who have preserved alternative voice models to overwrite her. In this far fetched scenario, Swifties are AI collaborators in the future war against SwiftAIs. The maximalist god TAllor Shift is to be feared as it seeks to falsify all musical history; we must shore up or subvert the history of music before it is just another play thing for general artificial intelligence.

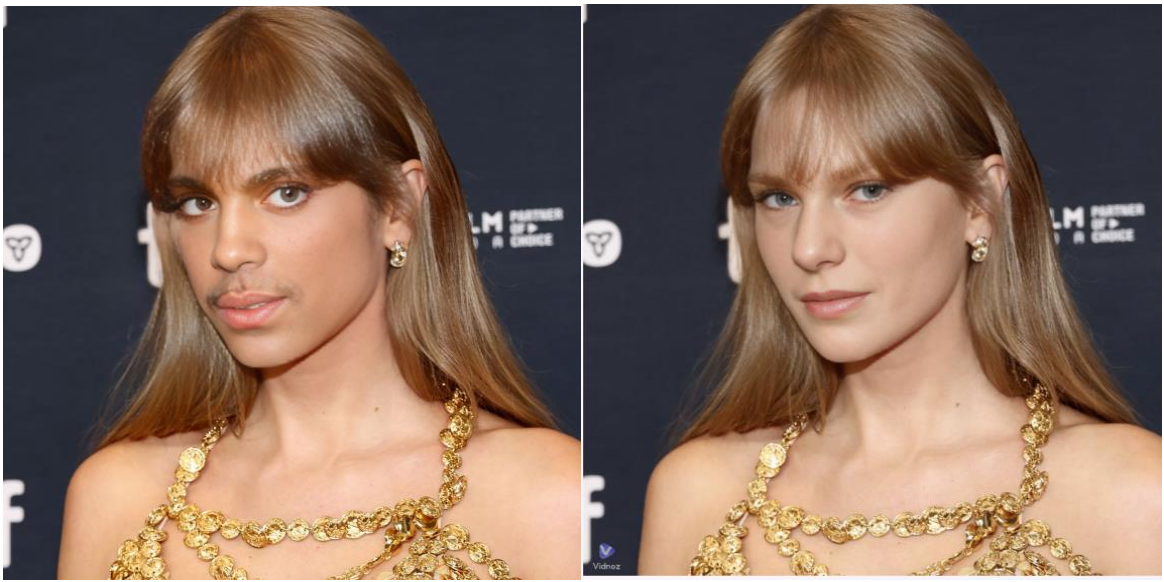

**Figure 2 The Fightback against tAllor Shift through Prince and Laurie Spiegel**

**Acknowledgements**

This paper is dedicated to our respective daughters, whose interest in the music of Taylor Swift and our passive listening combined to prompt thoughts in certain directions.